\documentclass[twocolumn,prl,final,aps]{revtex4}
\usepackage{amsmath,amssymb,amsthm}
\usepackage{pstricks,pst-plot}
\usepackage{subfigure}
\usepackage{times}

\newcommand{\comment}[1]{}

\newcommand{\ket}[1]{|#1\rangle}

\newcommand{\CTRL}[0]{{\operatorname{\textsc{ctrl}}}}
\newcommand{\SIFT}[0]{{\operatorname{\textsc{sift}}}}
\newcommand{\TEST}[0]{{\operatorname{\textsc{test}}}}
\newcommand{\INFO}[0]{{\operatorname{\textsc{info}}}}
\newcommand{\final}[0]{{\operatorname{final}}}

\newtheorem{prop}{Proposition}

\newtheorem{corol}{Corollary}
\newtheorem{theorem}{Theorem}
\theoremstyle{remark}

\theoremstyle{definition}

\begin{document}

\title{Quantum Key Distribution with Classical Bob}

\author{Michel Boyer$^{1}$, Dan Kenigsberg$^2$ and Tal Mor$^2$\\
\small 1. D\'epartement IRO, Universit\'e de Montr\'eal
  Montr\'eal (Qu\'ebec) H3C 3J7 \textsc{Canada} \\
\small 2. Computer Science Department, Technion, 
  Haifa 32000 \textsc{Israel}}

\date{\today}
\begin{abstract}

Secure key distribution among two remote 
parties is impossible when
both are classical, unless some unproven (and arguably unrealistic) 
computation-complexity assumptions are made,
such as the difficulty of factorizing large numbers.
On the other hand, %
a secure key distribution \emph{is} possible when both
parties are quantum.
What \emph{is} possible when only one party (Alice) is quantum, yet the other
(Bob) has only classical capabilities?
We present a protocol
with this constraint,
and 
prove its robustness against attacks: we prove
that any attempt of an adversary to obtain information
(and even a tiny amount of  
information)  
 necessarily induces
some errors that the legitimate users could notice.

\end{abstract}

\maketitle
\paragraph{Introduction.}
\label{sec:intro}

Processing information using quantum two-level systems (qubits),
instead of
classical two-state systems (bits), has lead to many striking results
such as the teleportation of unknown quantum states
and quantum algorithms
that are exponentially faster than their known classical counterpart.
Given a quantum computer,  Shor's factoring algorithm would render many of the currently
used encryption protocols completely insecure, 
but as a countermeasure, quantum information processing has also
given
quantum cryptography. 
Quantum key distribution
was invented by Bennett and Brassard (BB84),
to provide a new type of solution to
one of the most important cryptographic problems: the transmission
of secret messages.
A key distributed via quantum cryptography
techniques can be secure even against an eavesdropper with unlimited
computing power,
and the security is %
guaranteed forever.

The conventional setting is as follows:
Alice and Bob have labs that are perfectly secure,
they use qubits for their quantum communication, and
they have access to a classical communication channel
which can be heard, but cannot be jammed
(i.e.\ cannot be tampered with) by the eavesdropper.
The last assumption can easily be justified if Alice
and Bob can broadcast messages, or if they already share some
small number of secret bits in advance, to authenticate the
classical channel.

%
%
In the well-known BB84 protocol as well as in all 
other suggested protocols, both Alice and Bob perform quantum 
operations on their qubits (or on their quantum systems). 
Here we 
present, for the first time, a protocol in which one party 
(Bob) is classical. 
For our purposes, 
any two orthogonal states of the quantum two-level system can be chosen  
to be the computational basis $\ket{0}$ and $\ket{1}$.
For reasons that will soon become clear, 
we shall now call the computational basis ``classical'' and we shall use the 
classical notations $\{0,1\}$ to describe  
the two quantum states  
$\{\ket{0},\ket{1}\}$    
defining this basis.
In the protocol we present, a quantum channel travels from Alice's lab to
the outside world and back to her lab.
Bob can access a segment of the channel, and whenever a qubit 
passes through that segment Bob can either 
let it go undisturbed or %
(1).--- measure the qubit in the classical $\{0,1\}$ basis, and
(2).--- prepare a (fresh) qubit in the classical basis, and send it.
If \emph{all} parties were limited to performing only 
operations (1) and (2), or doing nothing, 
they would always be working with qubits in the classical basis, 
and could never obtain any
quantum superposition of the computational-basis states; the qubits can then
be considered ``classical bits''; the resulting protocol would then be
equivalent to a \emph{fully} classical protocol, and therefore, the operations
themselves shall here be considered classical.
We thus term this protocol ``QKD protocol with classical Bob''. One might
use the name 
Semi-Quantum Key Distribution (SQKD), 
since only one party performs operations
beyond the above. 

The question of how ``quantum'' a protocol should be in order to achieve 
a significant 
advantage over all classical protocols is of great interest.
For example, \cite{Popescu99,JL02,BBKM02,KMR06} discuss whether entanglement is
necessary for quantum computation, \cite{nonloc.noent99} shows nonlocality
without entanglement, and \cite{GPW05,FuchsSasaki03} discuss how
much of the information carried by various 
quantum states is actually classical.
We extend this discussion into another domain: 
{\em quantum cryptography}.
Such partially-quantum or semi-quantum protocols of various types 
might even have advantages over 
fully quantum protocols, if they are easier to implement in practice.
For instance, NMR quantum computing is among the most
successful implementations of quantum computing devices while 
the performed NMR experiments were proven to use 
no entanglement~\cite{Popescu99}.
Whether SQKD could also have potential practical advantages or not  
is left for future research.

%
To define our protocol we 
follow the definition (see for instance~\cite{BBBMR06}) 
of the most standard QKD protocol, BB84.
The BB84 protocol consists of two major parts: a first part that is aimed at
creating a \emph{sifted key}, and a second (fully classical) 
part aimed at extracting an error-free,
secure, \emph{final key} from the sifted key. 
In the first part of BB84, Alice randomly selects a binary value and
randomly selects in which basis to send it to Bob, either 
the computational (``$Z$'') basis $\{\ket0,\ket1\}$, or 
the Hadamard (``$X$'') basis $\{\ket+,\ket-\}$.
Bob measures each qubit in either basis at random. 
{\em An equivalent description is obtained if Alice
and  Bob use only the classical operations (1) and (2) above
and the Hadamard}~\footnote{$H\ket{0}=\ket{+}$;
$H\ket{1}=\ket{-}$.} {\em quantum gate $H$.}
After all qubits have been
sent and measured, Alice and Bob publish which bases they used. For
approximately half of the qubits Alice and Bob used mismatching bases and these
qubits are discarded. The values of the rest of the bits make the sifted key.
The sifted key is identical in Alice's and Bob's hands if  
the protocol is error-free and if there is no eavesdropper (known as Eve)
trying to learn the shared bits or some function of them.
In the second part Alice and Bob use some of the bits of the sifted key 
(the $\TEST$ bits) to test the error-rate, 
and if it is below some pre-agreed threshold, 
they select an $\INFO$ string from the rest of the sifted key. 
Finally, 
an error
correcting code (ECC) is used to correct the errors on the $\INFO$ string
(the $\INFO$ bits), and privacy
amplification (PA)
is used to derive a shorter but unconditionally
secure final key from these $\INFO$ bits.
At that point we would like to mention a key feature relevant
to our protocol:
it is sufficient to use qubits in just one basis, $Z$, 
for generating the $\INFO$ string,
while the other basis is used only for finding the actions of
an adversary~\cite{Mor98}.

A conventional measure of security is the information
Eve can obtain on the final key,
and a security proof usually calculates (or puts bounds on)
this information. 
The strongest (most general) attacks allowed by quantum mechanics
are called {\em joint attacks}. 
These attacks are aimed to learn something about the final 
(secret) key directly, 
by using a probe through which all qubits pass, and by measuring
the probe after all classical information becomes public.
Security against all joint attacks is considered as ``unconditional security''.
The security of BB84 (with perfect qubits
sent from Alice to Bob) against all joint attacks was first proven 
in~\cite{Mayers,Shor-Preskill,BBBMR06}   
via various techniques.
 
\paragraph{Robustness.}
An important step in studying security is a proof of robustness; see for
instance \cite{BBM92}
for robustness proof of their entanglement-based protocol,
and \cite{SARG04} 
for suggesting a protocol secure against the photon-number-splitting 
(PNS) attack,
and for proving its robustness.
Robustness of a protocol means that any adversarial attempt
to learn some information on the key  
necessarily induces some disturbance. 
It is a special case, in zero noise, of
the more general ``information versus disturbance" measure which provides
explicit bound on the information available to Eve as a function of the induced
error. 
Robustness also generalizes the no-cloning theorem: while the 
no-cloning theorem states that a state cannot be cloned, robustness means that 
any attempt to make an imprint of a state (even an extremely weak 
imprint) necessarily disturbs the quantum state.

\begin{description}
\item
Definitions:
A protocol is said to be
\emph{completely robust} if nonzero information acquired by Eve on the 
$\INFO$ string 
(before Alice and Bob perform the ECC step)
implies nonzero
probability that the legitimate participants find errors on the bits tested by the protocol.
A protocol is said to be \emph{completely nonrobust} if Eve can 
acquire the $\INFO$ string without inducing any error on the bits tested by the protocol.
A protocol is said to be
\emph{partly robust} if Eve can acquire some limited information on the 
$\INFO$ string without inducing any error on the bits tested by the protocol.
\end{description}

\begin{figure}
  \subfigure[~]{
  \begin{pspicture}(0,-0.5)(3.5,2.5)
    \psline[linearc=1,showpoints=false]{-}(0,0)(1,0.3)(2,1.6)(3,1.9)
    \psline[linearc=1,linestyle=dashed,linecolor=gray,arrowlength=3]{-}(0,1)(2,1.2)(3,1.8)
    \psline[linestyle=dotted,dotsep=1pt](0,2)(3,2)
    \psaxes[labels=none,ticks=y,dy=2,tickstyle=top]{->}(3,2.5)
    \rput{L}(-0.2,1.5){Info. on $\INFO$ bits}
    \rput{U}(1.5,-0.2){Dist. to tested bits}
  \end{pspicture}
  }
  \subfigure[~]{
  \begin{pspicture}(0,-0.5)(3,2.5)
    \pscurve[showpoints=false,linestyle=dotted]{-}%
            (1.1,0)(2,1.7)(3,2.0)
    \psaxes[labels=none,ticks=y,dy=2,tickstyle=top]{->}(3,2.5)
    \rput{L}(-0.2,1.5){Info. on final key}
    \rput{U}(1.5,-0.2){Dist. to tested bits}
    \rput{U}(0.55,0.25){$\overbrace{~~~~~~~~~}^{\mathrm{threshold}}$}
  \end{pspicture}
  }
  
  \caption{\footnotesize(a) Eve's maximum (over all attacks) information 
  on the $\INFO$ string vs. the allowed disturbance on the bits tested by
  Alice and Bob, in a completely robust (solid line), partly robust
  (dashed), and completely nonrobust (densely dotted) protocol.
  (b) Robustness should not be confused with security; Eve's maximum 
  information on the \emph{final} key vs. allowed disturbance in a secure
protocol; such a protocol could be completely or partly robust.}
  \label{fig:robustness}
  \end{figure}
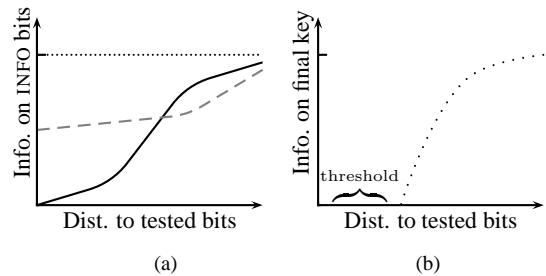

Partly-robust protocols could still be secure, yet completely nonrobust 
protocols are automatically proven insecure 
(Cf.\ Fig.~\ref{fig:robustness}).
As one example, BB84 is fully robust when qubits are used by Alice and Bob 
but it is only partly robust 
if photon pulses are used and sometimes two-photon pulses
are sent. 
%
%

Here we prove that our protocol for ``quantum key distribution 
with classical Bob'' is completely robust. Another protocol and a proof of its
robustness are omitted for the sake of
brevity, and will be provided in a future work.

\paragraph{A mock protocol and its complete nonrobustness.}
\label{sec:idea}

Consider the following mock protocol:
Alice generates a random qubit in the $Z$-basis.
She chooses randomly whether to do nothing,
or apply Hadamard gate to transform the qubit to 
the $X$-basis.
Bob flips a coin to decide whether 
to measure Alice's qubit in the $Z$-basis 
(to ``$\SIFT$'' it) or to reflect it back 
(``$\CTRL$''), without causing any
modification to the information carrier.
In case Alice chose $Z$ and Bob decided to $\SIFT$, i.e.\ to measure in the $Z$
basis, they share a random
bit that we call $\SIFT$ bit (that may, or may not, be confidential). 
In case Bob chose $\CTRL$, Alice can check if the qubit returned
unchanged, by measuring it in the basis she sent it. 
In case Bob chose to $\SIFT$ and Alice chose the $X$ basis,
they discard that bit.
The above iteration is repeated for a predefined number of times. 
At the end of the quantum part of the protocol Alice and Bob share, 
with high probability, a considerable amount of $\SIFT$ bits (also known
as the ``sifted key'').
In order to make sure that Eve cannot gain much information by measuring
(and resending) all
qubits in the $Z$ basis, 
Alice can check whether they have a low-enough level of discrepancy
on the $X$-basis $\CTRL$ bits.
In order to make sure that their sifted key is reliable, 
Alice and Bob must sacrifice
a random subset of the $\SIFT$ bits, which we denote as $\TEST$ bits, 
and remain with a string of bits which we call $\INFO$ bits 
($\INFO$ and $\TEST$ are common in QKD, e.g., in BB84 as previously described).

By comparing the value of the $\TEST$ bits,
Alice and Bob can estimate the error rate on the $\INFO$ bits.
If the error rate on the $\INFO$ bits 
is sufficiently small, 
they use an appropriate Error Correction Code (ECC) 
in order to correct the errors.
If the error rate on the $X$-basis $\CTRL$ bits  
is sufficiently small, Alice and Bob can bound Eve's information,
and use an appropriate Privacy
Amplification (PA) in order to 
obtain any desired level of privacy.

At first glance, 
this protocol may look like a nice way to transfer 
a secret bit from quantum Alice to classical Bob: It is
probably resistant to opaque (intercept-resend) attacks,
and probably also against all collective attacks (where Eve uses a different
probe in each access to each qubit). 
However, it is {\em completely non-robust}; 
Eve could learn all bits of the $\INFO$ string
using a trivial %
attack that induces no error on the bits tested by Alice 
and Bob (the $\TEST$ and
$\CTRL$ bits).
She would not measure the incoming qubit, but rather perform a cNOT from 
it into a $\ket{0^E}$ ancilla~\footnote{By ``cNot from
$A$ into $B$'' we mean that $A$ is the control qubit and $B$ is the
target, as is commonly called.}. 
If Alice chose $Z$ and Bob decide to $\SIFT$ (i.e.\ measures in the $Z$-basis), 
she measures her ancilla and
obtains an exact copy of their common bit, thus inducing no
error on $\TEST$ bits and learning the $\INFO$ string.
If, however, Bob decides on $\CTRL$, i.e.\ reflects the qubit,
Eve would perform another cNOT from the returning qubit into her ancilla. 
This would reset her ancilla, erase the interaction she performed, and
induce no error on $\CTRL$ bits, thus
removing any chance of her being caught. 

%
Note
that in this mock protocol, Bob did not use classical operation~(2) at all. 
In the following section we present a protocol in which Bob 
always sends a qubit to Alice (making use of operation~(2) when needed).
By always returning all qubits he enforces Eve to delete any information she
gained, or else some error is potentially induced.

\paragraph{A Semi-Quantum Key Distribution Protocol.}
%
The following protocol remedies the above weakness
by not letting Eve know which is
a $\SIFT$ qubit (that can be safely measured in the computational basis)
and which is a $\CTRL$  qubit (that should be returned to Alice unchanged).
The protocol is aimed at creating an $n$-bit $\INFO$ string to be
used as the seed for an $m$-bit shared secret key.

Let the integer $n$ be the desired length of the $\INFO$ string,  
and
let $\delta>0$ be
some fixed parameter.
\begin{enumerate}
\item Alice generates $N = 8n(1+\delta)$ random qubits in the $Z$ basis.
For each of the qubits,
she randomly selects whether to apply the Hadamard gate (``$X$'') or do nothing (``$Z$''). 

\item For each qubit arriving,
Bob chooses randomly either to reflect it ($\CTRL$)
or to measure it in the $Z$ basis  
and resend it in the same state he found  (to $\SIFT$ it).
Bob sends the first qubit to 
Alice after receiving the last qubit, in the same order he received them.
\label{step:bob-choice}
\item Alice measures each 
qubit in the basis she sent it.
\label{step:alice-collect}
\item Alice publishes which were her $Z$ bits and Bob publishes
which ones he chose to $\SIFT$.
\label{step:alice-compare}
\end{enumerate}
It is expected that for approximately $N/4$ bits, Alice used the $Z$ basis
for transmitting, and Bob chose to $\SIFT$;
these are the $\SIFT$ bits, which form the sifted key.
For approximately $N/4$ bits,
Alice used the $Z$ basis
and Bob chose $\CTRL$; we refer to these bits as $Z$-$\CTRL$.
For approximately $N/4$ bits,
Alice used the $X$ basis
and Bob chose $\CTRL$; we refer to these bits as $X$-$\CTRL$.
The rest of the bits (those sent in the $X$ basis 
but chosen as $\SIFT$ by Bob)
are ignored.
\begin{enumerate}
\setcounter{enumi}{4}
\item
Alice checks the error-rate on the $\CTRL$ bits and if either
the $X$ error-rate or the $Z$ error-rate is higher than some
predefined threshold $P_{\CTRL}$ the protocol aborts.
\item Alice chooses
at random $n$ $\SIFT$ bits to be $\TEST$ bits.
She publishes which are the chosen bits.
Bob publishes the value of these $\TEST$ bits. 
Alice checks the error-rate on the $\TEST$ bits and if it is higher than some
predefined threshold $P_{\TEST}$ the protocol aborts.
\end{enumerate}
The protocol aborts if there are not
enough bits to perform Step~6 or Step~7;
this happens with exponentially small probability.
\begin{enumerate}
\setcounter{enumi}{6}
\item
Alice and Bob select the first $n$ remaining $\SIFT$ bits to be used
as $\INFO$ bits.
\item Alice publishes ECC \& PA data;
she and Bob use them to extract the $m$-bit final key from the $n$-bit $\INFO$ string.
\end{enumerate}

\paragraph{A Proof of Robustness.}

We show that Eve cannot obtain information on $\INFO$ bits
without being detectable. 

\label{sec:robust}

\paragraph*{Modeling the protocol.} Each time the protocol is executed,
Alice sends to Bob a state $\ket{\phi}$ which is a product of $N$ qubits, each of which
is either $\ket{+}$, $\ket{-}$, $\ket{0}$ or $\ket{1}$; 
those qubits are indexed %
 from $1$ to $N$. %
Each of them is either measured
by Bob in the $Z$ basis and resent as it was measured, or simply reflected. 
Let $m=\{m_1,m_2 ... m_r\}$ a set of $r<N$ integers
$1 \leq m_1 < m_2 ... < m_r \leq N$, describing the qubits chosen by Bob
as $\SIFT$. 
For $i\in \{0,1\}^N$, we denote
$ %
i_m = i_{m_1}i_{m_2}\ldots i_{m_r}
$ %
 the substring of $i$ of length $r$ selected by
the positions in $m$;
of course $\ket{i_m} = \ket{ i_{m_1}i_{m_2}\ldots i_{m_r}}$.

In the protocol, it is assumed that Bob has no quantum register; he measures the
 qubits as they come in. The physics
would however be exactly the same if Bob used a quantum register of $r$ qubits initialized
in state $\ket{0^B} = \ket{0^r}$ ($r$ qubits equal to $0$), applied the unitary transform
defined by
\(
U_m \ket{i}\ket{0^B} = \ket{i}\ket{i_m}
\)
for $i\in\{0,1\}^N$, sent back $\ket{i}$ to Alice
 and postponed his measurement to be performed on that quantum register $\ket{i_m}$;
 the qubits indexed by $m$ in $\ket{i}$ are thus automatically both measured and
 resent, and those not in $m$
 simply reflected;
 the $k$th qubit sent by Alice is a $\SIFT$ bit if $k\in m$ and is either $\ket{0}$ or $\ket{1}$;
it is a $\CTRL$ bit if $k\notin m$.
This physically-equivalent modified protocol
 simplifies the analysis, and we shall thus model Bob's measurement and resending, or
 reflection, with $U_m$. 
\paragraph*{Eve's attack.}
Eve's  most general attack is comprised of two unitaries: $U_E$ attacking qubits as they
go from Alice to Bob and  $U_F$ as they go back from Bob to Alice, where $U_E$ and
$U_F$ share a common probe space with initial state $\ket{0^E}$.
The shared probe allows Eve to make the attack on the returning qubits depend
on knowledge acquired by $U_E$ (if Eve does not take advantage of that fact,
then the ``shared probe'' can simply be the composite system comprised of two 
independent probes). Any attack where Eve would make $U_F$ depend
on a measurement made after applying $U_E$ can be implemented by unitaries
$U_E$ and $U_F$ 
with controlled gates
so as to postpone measurements; since we are giving Eve all the power of quantum
mechanics, the difficulty of building such a circuit is of no concern.

\paragraph*{The final global state.} 
Delaying all measurements allows considering the final global state
of the Eve+Alice+Bob system before all measurements. 
To state $\ket{\phi}$ sent by Alice, Eve attaches
the probe $\ket{0^E}$, applies $U_E$ to $\ket{0^E}\ket{\phi}$ and sends
Bob his part of the system, $N$ qubits. Taking into account Bob's probe
$\ket{0^B}$, the global state is now
$[U_E\otimes I_M]\ket{0^E}\ket{\phi}\ket{0^B}$ where $I_M$ is the identity
on Bob's probe space. Then, Bob applies $U_m$ to his part of the system,
which corresponds to applying $I_E\otimes U_m$ to the previous global
state where $I_E$ is the identity on Eve's probe space. Eve's attack
on the returning qubits corresponds to applying the 
unitary $U_F\otimes I_M$ and the final global state is
\begin{equation}\label{finalglobal}
[U_F\otimes I_M][I_E\otimes U_m][U_E\otimes I_M]\ \ket{0^E}\ket{\phi}\ket{0^B}.
\end{equation}

\begin{prop}\label{propositiondefeideffi}
If  $U_E$ induces no error on $\TEST$ bits, then there are
states $\ket{E_i}$ in Eve's probe space such that for all $i\in\{0,1\}^N$
\begin{equation}\label{defei}
U_E\ket{0^E}\ket{i} = \ket{E_i}\ket{i}
\end{equation}
If, moreover, $(U_E,U_F)$ induces no error on $\CTRL$ bits, then there are states
$\ket{F_i}$ in Eve's probe space such that for all $i\in \{0,1\}^N$,
\begin{equation}\label{deffi}
U_F\ket{E_i}\ket{i} = \ket{F_i}\ket{i}.
\end{equation}
\end{prop}
\begin{proof}
When $U_E$ is applied onto the computational basis, 
$U_E\ket{0^E}\ket{i} = \sum_{j} \ket{E_{i,j}} \ket{j}$.
If for some index $k$ there is some $j$ such that $i_k\neq j_k$ and $\ket{E_{i,j}}\neq 0$,
then by choosing $m$ such that $k\in m$, Bob can detect this as an error on bit $k$.
For Eve's attack to be undetectable on $\TEST$ bits,
$U_E$ must thus be such that $U_E\ket{0^E}\ket{i} = \ket{E_{i,i}}\ket{i}$,
namely, $\ket{E_{i,j}} = 0$ for any $j \neq i$,
and $\ket{E_i} = \ket{E_{i,i}}$ satisfies Eq.~(\ref{defei}). 
If Alice sent state $\ket{i}$ for $i\in\{0,1\}^N$, the global
state is then $\ket{E_i}\ket{i}\ket{i_m}$
and $U_F\ket{E_{i}}\ket{i} = \sum_{j} \ket{F_{i,j}} \ket{j}$. 
In order for Eve's attack to be undetectable on  
$Z$-$\CTRL$ bits (whose index is not in $m$),  
$U_F$ must be such that $U_F\ket{E_{i}}\ket{i} = \ket{F_{i,i}}\ket{i}$,
namely, $\ket{F_{i,j}} = 0$ for any $j \neq i$ and $\ket{F_i} = \ket{F_{i,i}}$
then satisfies Eq.~(\ref{deffi}).%
\end{proof}
\begin{corol} If the attack $(U_E,U_F)$ induces no error on $\TEST$ and $\CTRL$ bits, then
(for all $i\in\{0,1\}^N$ and all $m$) 
the final global state (\ref{finalglobal}) 
if $\ket{\phi} = \ket{i}$ is
\begin{equation}\label{summerize0}
\ket{F_{i}}\ket{i}\ket{i_m}.\end{equation}
\end{corol}
We now show that if Eve's attack is undetectable by Alice and Bob, then
Eve's final state $\ket{F_i}$ 
is independent
of the string $i\in \{0,1\}^N$. More precisely
\begin{prop}\label{propprot2}
If  $(U_E, U_F)$ is an attack that induces no error on
 $\TEST$ and $\CTRL$ bits, 
 and if $\ket{F_i}$ is given by Eq.~(\ref{summerize0}), then %
 for all $i, i' \in \{0,1\}^N$
 \begin{equation}\label{property0}
 i,i'\in \{0,1\}^N \quad \implies \quad \ket{F_i} = \ket{F_{i'}}
\ .
 \end{equation}
\end{prop}

\begin{proof}
Eq.~(\ref{property0}) means that any of the $N$ bits of $i\in\{0,1\}^N$ can be flipped at will without 
affecting Eve's final state
$\ket{F_i}$. We thus need only prove that for any two bit strings $i, i'\in \{0,1\}^N$ that differ only on one
bit, say bit $k$,
the equality $\ket{F_i} = \ket{F_{i'}}$ holds. We assume wlg that $i_k=0$
and $i'_k=1$. If Alice chooses qubit $k$ to be $X$-$\CTRL$ and
chooses all the other qubits
to be those of $i$ and $i'$, then this means that the state $\ket{\phi}$ she sends is
$\frac{1}{\sqrt{2}}\left[\ket{i} + \ket{i'}\right]$. %
Assume now that
Bob reflects bit $k$, i.e.\ that $k\notin m$. This implies that $i_m= i'_m$.
By Eq.~(\ref{summerize0}) and linearity,
the final state is $\frac{1}{\sqrt{2}}\left[ \ket{F_i}\ket{i} + \ket{F_{i'}}\ket{i'}\right]\ket{i_m}$.
Since we are interested only in Alice's $k$th qubit, we trace-out all the other qubits
in Alice and Bob's hands. The resulting state 
\begin{equation}\label{sttraced}
\frac{1}{\sqrt{2}}\left[\ket{F_i}\ket{0} + \ket{F_{i'}}\ket{1} \right]
\end{equation}
must be such that the probability of Alice measuring $\ket{-}$ is $0$. Replacing
$\ket{0}$ and $\ket{1}$ by their value in terms of $\ket{+}$ and $\ket{-}$,  state (\ref{sttraced})
rewrites as
$ %
\frac{1}{2}\Big[\ket{F_i}+\ket{F_{i'}}\Big] \ket{+} + \frac{1}{2}\Big[\ket{F_i}-\ket{F_{i'}}\Big]\ket{-}
$ %
and the probability of measuring $\ket{-}$ is $0$
iff $\frac{1}{2}\Big[\ket{F_i}-\ket{F_{i'}}\Big]=0$ i.e.\ 
$\ket{F_i} = \ket{F_{i'}}$.
\end{proof}

\begin{theorem} The protocol is completely robust:
for any attack $(U_E, U_F)$ %
inducing no error
on $\TEST$ and $\CTRL$ bits, Eve's final state is independent of the states
$\ket{\phi}$ sent by Alice, and Eve is thus left with no information on the $\INFO$ string.
\end{theorem}
\begin{proof}
By Proposition~\ref{propprot2}, there is a state $\ket{F_\final}$ in Eve's probe space
s.t.\ for all $i\in \{0,1\}^N$, Eve's final state
$\ket{F_i} = \ket{F_\final}$.
If Alice sends any superposition
$\ket{\phi} = \sum_{i}  c_i \ket{i}$ and Bob chooses any set $m$ of bits to be 
measured (leaving at least one $\CTRL$ bit). Using 
Eq.~(\ref{summerize0}) with $\ket{F_i} = \ket{F_\final}$ for
all $i$ and linearity gives 
 $ %
\ket{F_\final} \sum_i c_i\ket{i}\ket{i_m}
$  %
as the final global state of the system; %
Eve's probe state  $\ket{F_\final}$ is independent of %
$i_m$ and therefore of the $\SIFT$ and $\INFO$ bits. %
\end{proof}

\paragraph{Conclusion.}
We presented a protocol for QKD with one party who performs only classical
operations
and proved its robustness.
We believe that our work sheds light on how much ``quantumness'' is 
required in order to perform classically-impossible tasks in general, 
and secret key distribution in particular.
This work was partially supported by the Israeli MOD.
We thank Moshe Nazarathy for providing the motivation for this research.




\end{document}